\input{aipcheck}

\documentclass[
    ,final            
  ]
  {aipproc}

\layoutstyle{8x11single}

\usepackage{color}

\begin{document}

\title{The AMY (Air Microwave Yield) experiment to measure the GHz emission from air shower plasma}

\classification{96.50.sd}
\keywords      {UHECR, microwave emission, test beam}

\author{J. Alvarez-Mu\~niz}{
  address={Depto. de Fisica de Particulas, Universidad de Santiago de Compostela, Santiago de Compostela, Spain}
}

\author{M. Blanco}{
  address={Laboratoire de Physique Nucl\'eaire et de Hautes Energies (LPNHE), Universit\'es Paris 6 et Paris 7, CNRS-IN2P3, Paris, France}
}

\author{M. Boh\'a\v cov\'a}{
  address={Institute of Physics, Academy of Sciences of the Czech Republic, Prague, Czech Republic}
}

\author{B. Buonomo}{
  address={Istituto Nazionale di Fisica Nucleare - Laboratori Nazionali di Frascati, Via E. Fermi, 40 - 00044 Frascati, Italy}
}

\author{G. Cataldi}{
  address={Sezione INFN, Lecce, Italy}
}

\author{M. R. Coluccia}{
  address={Dipartimento di Matematica e Fisica Ennio De Giorgi, Universit\`{a} del Salento, Lecce, Italy}
 ,altaddress={Sezione INFN, Lecce, Italy}
}

\author{P. Creti}{
  address={Sezione INFN, Lecce, Italy}
}

\author{I. De Mitri}{
  address={Dipartimento di Matematica e Fisica Ennio De Giorgi, Universit\`{a} del Salento, Lecce, Italy}
 ,altaddress={Sezione INFN, Lecce, Italy}
}

\author{C. Di Giulio}{
  address={Dipartimento di Fisica, Universit\`{a} di Roma Tor Vergata , Roma, Italy}
 ,altaddress={Sezione INFN, Roma Tor Vergata, Italy}
}

\author{P. Facal San Luis}{
  address={University of Chicago, Enrico Fermi Institute Kavli Institute for Cosmological Physics, Chicago, USA }
}

\author{L. Foggetta}{
  address={Istituto Nazionale di Fisica Nucleare - Laboratori Nazionali di Frascati, Via E. Fermi, 40 - 00044 Frascati, Italy}
}

\author{R. Ga\"ior}{
  address={Laboratoire de Physique Nucl\'eaire et de Hautes Energies (LPNHE), Universit\'es Paris 6 et Paris 7, CNRS-IN2P3, Paris, France}
}

\author{D. Garcia-Fernandez}{
  address={Depto. de Fisica de Particulas, Universidad de Santiago de Compostela, Santiago de Compostela, Spain}
}

\author{M. Iarlori}{
  address={Dipartimento di Fisica, Universit\`{a} dell'Aquila and sezione INFN, l'Aquila, Italy}
}

\author{S. Le Coz}{
  address={Laboratoire de Physique Subatomique et de Cosmologie (LPSC), Universit\'e J. Fourier Grenoble,CNRS-IN2P3, Grenoble, France}
}

\author{A. Letessier-Selvon}{
  address={Laboratoire de Physique Nucl\'eaire et de Hautes Energies (LPNHE), Universit\'es Paris 6 et Paris 7, CNRS-IN2P3, Paris, France}
}

\author{ K. Louedec}{
  address={Laboratoire de Physique Subatomique et de Cosmologie (LPSC), Universit\'e J. Fourier Grenoble,CNRS-IN2P3, Grenoble, France}
}

\author{I. C. Mari\c{s}}{
  address={Universidad de Granada and C.A.F.P.E., Granada, Spain}
}

\author{D. Martello}{
  address={Dipartimento di Matematica e Fisica Ennio De Giorgi, Universit\`{a} del Salento, Lecce, Italy}
 ,altaddress={Sezione INFN, Lecce, Italy}
}

\author{G. Mazzitelli}{
  address={Istituto Nazionale di Fisica Nucleare - Laboratori Nazionali di Frascati, Via E. Fermi, 40 - 00044 Frascati, Italy}
}


\author{L. Perrone}{
  address={Dipartimento di Matematica e Fisica Ennio De Giorgi, Universit\`{a} del Salento, Lecce, Italy}
 ,altaddress={Sezione INFN, Lecce, Italy}
}


\author{S. Petrera}{
  address={Dipartimento di Fisica, Universit\`{a} dell'Aquila and sezione INFN, l'Aquila, Italy}
}

\author{P. Privitera}{
 address={University of Chicago, Enrico Fermi Institute Kavli Institute for Cosmological Physics, Chicago, USA}
}

\author{V. Rizi}{
  address={Dipartimento di Fisica, Universit\`{a} dell'Aquila and sezione INFN, l'Aquila, Italy}
}

\author{G. Rodriguez Fernandez}{
  address={Dipartimento di Fisica, Universit\`{a} di Roma Tor Vergata , Roma, Italy}
 ,altaddress={Sezione INFN, Roma Tor Vergata, Italy}
}

\author{F. Salamida}{
  address={ Institut de Physique Nucl\'{e}aire d'Orsay (IPNO), Universit\'{e} Paris 11, CNRS-IN2P3, France}
}

\author{G. Salina}{
  address={Sezione INFN, Roma Tor Vergata, Italy}
}

\author{M. Settimo}{
  address={Laboratoire de Physique Nucl\'eaire et de Hautes Energies (LPNHE), Universit\'es Paris 6 et Paris 7, CNRS-IN2P3, Paris, France}
}

\author{P. Valente}{
  address={Istituto Nazionale di Fisica Nucleare - Laboratori Nazionali di Frascati, Via E. Fermi, 40 - 00044 Frascati, Italy}
}

\author{ J. R. Vazquez}{
  address={Universidad Complutense de Madrid, Madrid, Spain}
}

\author{V. Verzi}{
  address={Sezione INFN, Roma Tor Vergata, Italy}
}

\author{C. Williams}{
 address={University of Chicago, Enrico Fermi Institute Kavli Institute for Cosmological Physics, Chicago, USA}
}

\begin{abstract}
The aim of the Air Microwave Yield (AMY) experiment is to investigate the Molecular Bremsstrahlung Radiation (MBR) emitted
from an electron beam induced air-shower. The measurements have been performed with a 510 MeV electron beam at the
Beam Test Facility (BTF) of Frascati INFN National Laboratories in a wide frequency range between 1 and 20 GHz. We
present the experimental apparatus and the first results of the measurements. Contrary to what have been reported in a previous
similar experiment~\cite{Gorham-SLAC}, we have found that the intensity of the emission is strongly influenced by the particular time structure
of the accelerator beam. This makes very difficult the interpretation of the emission process and a realistic extrapolation of
the emission yield to the plasma generated during the development of an atmospheric shower.
\end{abstract}

\maketitle


\section{Introduction}

The fluorescence technique is a well established method to detect the cosmic rays at energies > $10^{18}$ eV. It provides a calorimetric measurement of the primary energy from the detection of the radiation produced by the de-excitation of the atmospheric nitrogen. Fluorescence telescopes are used by the Pierre Auger Observatory~\cite{PAO} and Telescope Array~\cite{TARio}, which are the largest detectors of cosmic rays that were ever built. The fluorescence measurements are used to determine the energy scale~\cite{Verzi} of the surface array detector. However, this can be done only in small fraction of events detected by the surface array since the fluorescence telescopes can run only during moonless nights with an overall duty cycle of about 10\%.

Recently, a new detection technique alternative to the fluorescence one has been proposed~\cite{Gorham-SLAC}, in which cosmic rays are detected measuring the MBR radiation at frequencies of few GHz produced during the shower development.  MBR is expected to be emitted by low energy electrons during their interaction with the field of the neutral molecules.  Alike the fluorescence emission,  MBR should be isotropic and un-polarized. A GHz telescope would be very similar to a fluorescence one, but with the fundamental advantage of a 100\% duty cycle.

The new technique has been proposed after the first observation of the GHz radiation from air shower plasmas~\cite{Gorham-SLAC}. This was measured at the Stanford Linear Accelerator Center (SLAC) using a 28 GeV electron beam.
The measurements have been performed in a 1 m$^3$ copper anechoic Faraday chamber in the frequency band 1.5$-$6.0 GHz. Before entering into the chamber, the beam was collided with a 90\% Al$_2$O$_3$ $-$ 10\% SiO$_2$ target. The signal was found to decay exponentially with $\tau =$ 10 ns and to scale with the square of the beam energy, indicating that the emission is fully coherent. The measured density flux of the radiation was $4 \times 10^{-16} ~{\rm W/m^2 /Hz}$.

Due to the high potential of an MBR telescope, in the past few years several activities started trying to address the feasibility of this new detection technique. AMBER~\cite{Gorham-SLAC,Imen-Arena2014},
MIDAS~\cite{Imen-Arena2014,midas} and EASIER~\cite{Imen-Arena2014} are prototype GHz detectors installed at the Auger Observatory. Another detector, CROME, has been set up within the KASCADE-Grande
array~\cite{crome}. Several shower candidates have been detected, but the emission mechanism has not been identified. In general, the measurements disfavour the hypothesis of coherent emission, which is
in contrast to what has been reported in~\cite{Gorham-SLAC}.

Other measurements have been done using a 3 MeV electron beam of the Van de Graaff at the Argonne National Laboratory (USA)~\cite{maybe}  and using a 95 keV electron beam from an electrostatic gun~\cite{conti}. Both experiments reported a linear scaling of the radiation intensity with the beam energy, which means that the radiation is not emitted coherently. Moreover the measured intensity is significantly lower than the one reported in~\cite{Gorham-SLAC}.

It is worth noting that the inconsistencies between the recent measurements and the first test performed at SLAC~\cite{Gorham-SLAC} could be due to several factors. Although the MBR is a well known process in plasma physics, there are many uncertainties in the estimate of the emitted intensity~\cite{Gorham-SLAC,Imen-theory}. For example, the density of the plasma could affect significantly the degree of coherence of the radiation and in general, the plasma conditions could depend on the individual source of particles.

The aim of the AMY experiment is to study the MBR emission in a wide frequency range, between 1 and 20 GHz, using the electron beam of the BTF, Frascati. Contrary to the other experiments, the running conditions of AMY are rather similar to the ones of~\cite{Gorham-SLAC}. In this paper we present the experimental apparatus and the first results of the measurements.

\section{The experimental apparatus}

\begin{figure}[h]
\centering
\includegraphics[width=0.7\textwidth]{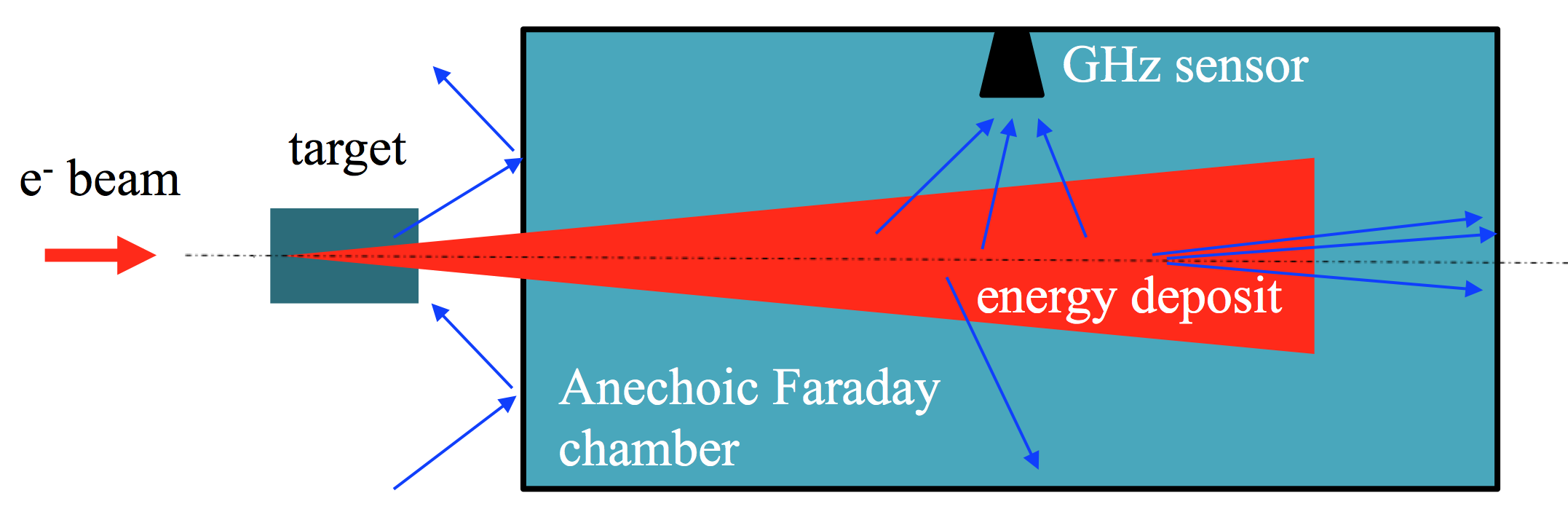} 
 \caption{Sketch of the AMY experiment.}
\label{Fig:AMYSketch}
\end{figure}

A sketch of the AMY experiment is shown in
Fig.~\ref{Fig:AMYSketch}. The radiation is observed with GHz sensors
within an anechoic Faraday chamber, which prevents the background and
reflected radiation produced in the BTF hall from entering into the
chamber. The beam collides with an interaction target of a
variable thickness. This allows to study the MBR radiation as a
function of the energy deposit inside the chamber. In fact, since the
MBR is expected to be produced by the secondary electrons, its
intensity should be proportional to the energy deposited as in the case of
the fluorescence radiation.

\subsection{The accelerator beam and the interaction target}

\begin{figure}[h]
\centering
\centerline{ \includegraphics[width=0.45\textwidth]{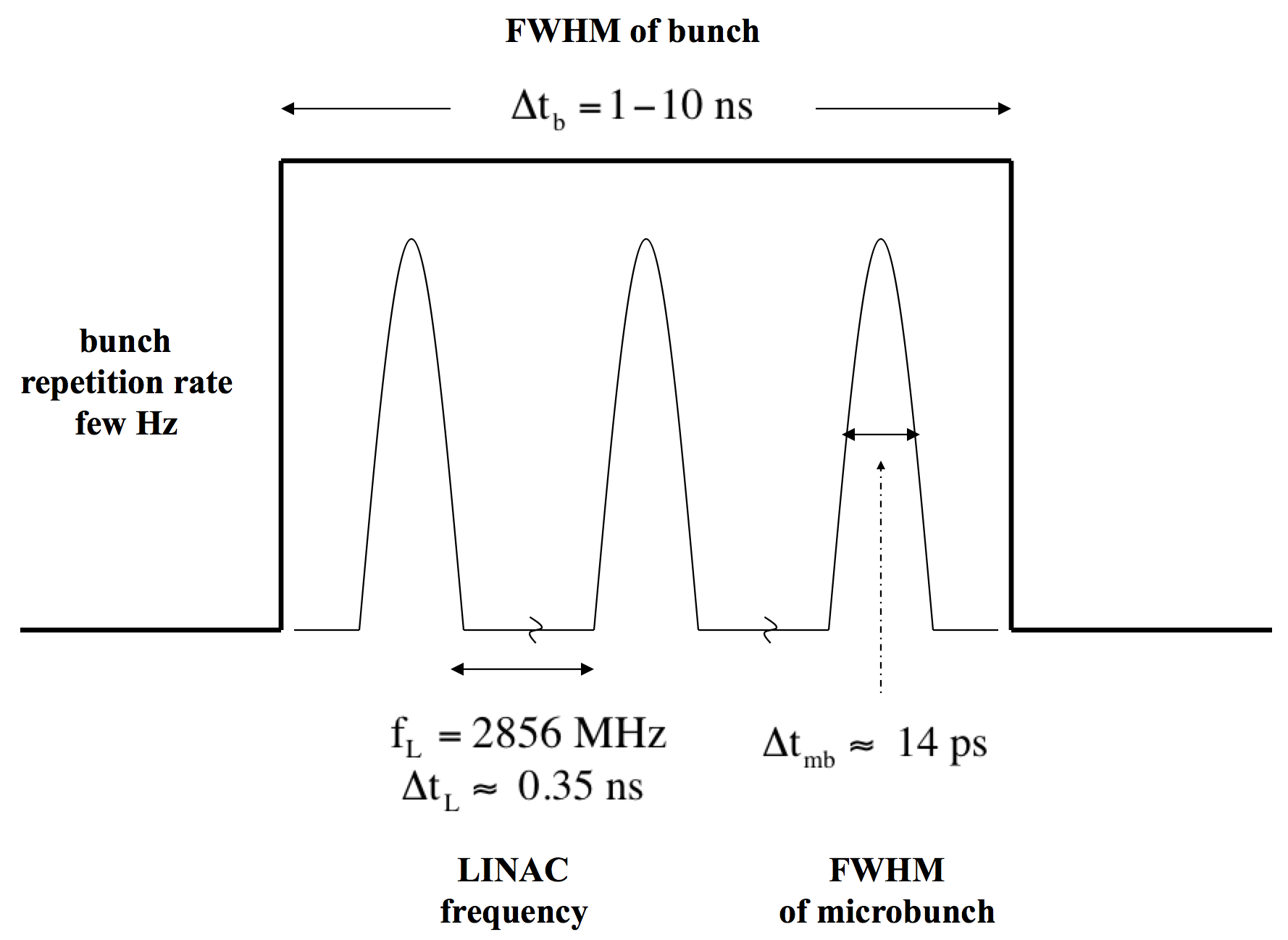} \includegraphics[width=0.4\textwidth]{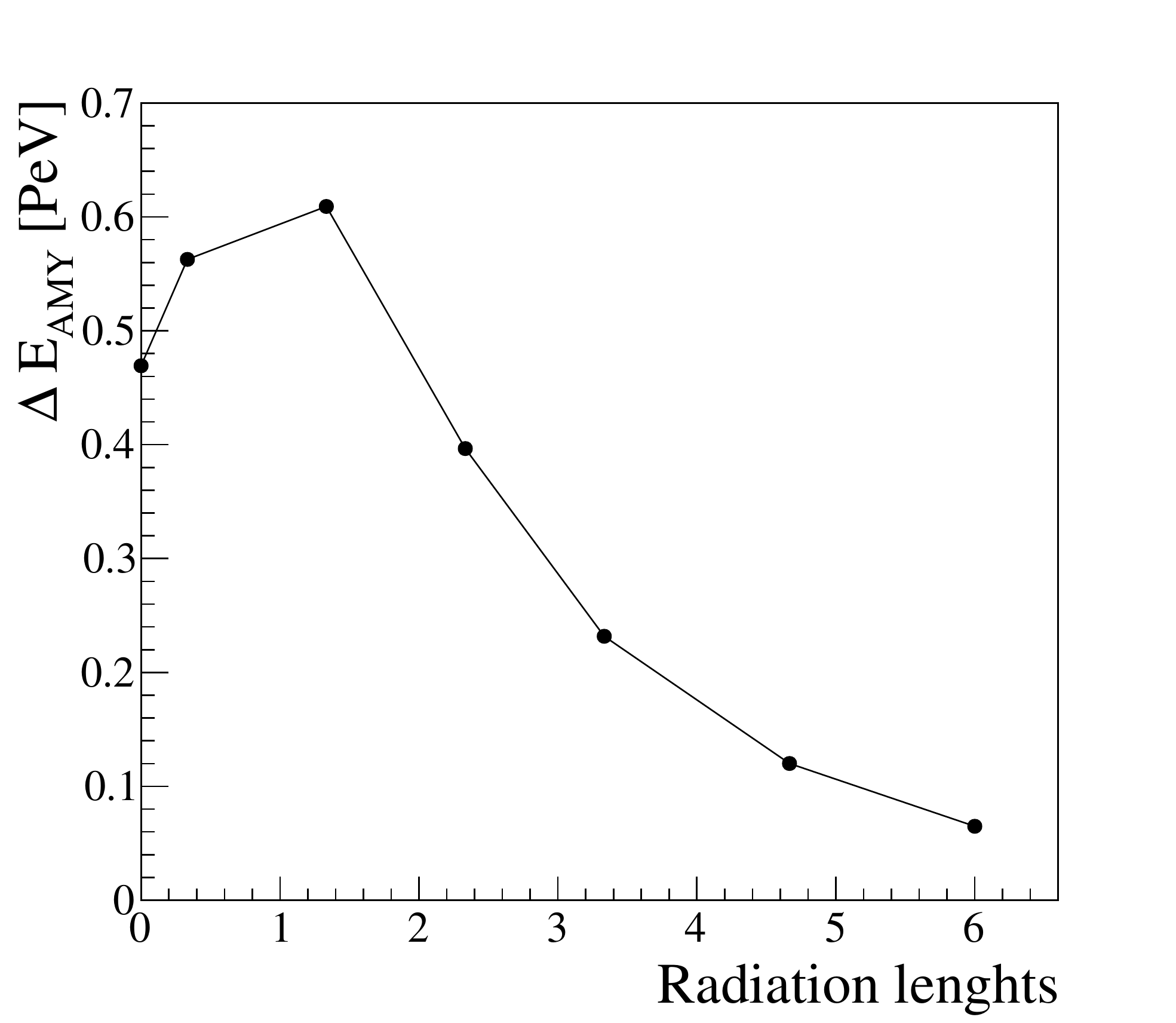} }
 \caption{Left panel: sketch of the time structure of the BTF beam. Right panel: simulated energy deposit within the AMY chamber.}
\label{Fig:BunchTarget}
\end{figure}

BTF~\cite{btf} is a part of the DA$\Phi$NE accelerator complex, which is composed of a dedicated transfer line, driven by a pulsed magnet, that allows to divert electrons or positrons from the end of the high intensity LINAC towards a 
100 m$^2$ experimental hall. The LINAC can provide electron bunches with a charge up to $10^{10}$ $e^-$/pulse, in an energy range between 25 and 750 MeV, with a bunch length between 1$-$10 ns and a maximum repetition rate of 50 Hz. The bunches accelerated by the LINAC are made of several microbunches with a FWHM of 14 ps separated by 0.35 ns, which corresponds to the inverse of the LINAC frequency 2.856 GHz (see the left panel of Fig.~\ref{Fig:BunchTarget}). 
The beam intensity is measured by an Integrating Current Transformer \emph{BERGOZ ICT 122-070-05-1} placed at the end of the beam pipe in the BTF hall.

The interaction target consists of 95\% pure Al$_2$O$_3$. It is made by six modules remotely controlled by a pneumatic system. Two modules have thickness of 10 cm, three of 7.5 cm and one of 2.5 cm. The maximum thickness is then 45 cm which corresponds to about 7 radiation lengths (X$_0$). The energy deposit within the chamber is simulated using the Geant4~\cite{geant4} software. The right panel of Fig. 2 shows the energy deposited by the typical beam used by AMY, $10^9$ electrons with energy 510 MeV. The energy deposit is shown for the various thickness of the target. We have also simulated the energy deposit in the SLAC test~\cite{Gorham-SLAC}. At the maximum, our energy deposit is about a factor 6 larger.

We have done three measurement campaigns, one in 2011 and two in 2012. The first two campaigns were very useful to study and improve the running conditions and the experimental apparatus. 
The bunch length was 10 ns, 3 ns and 1.5 ns. The latter was available only in the last test of 2012. 

\subsection{The anechoic Faraday chamber and the DAQ system}

The anechoic Faraday chamber was built at the mechanical shop of the physics department of the Roma Tor Vergata university. The dimensions of the chamber (2 m width $\times$ 4 m long) are mainly constrained by the space available in the BTF hall. To facilitate its \begin{figure}[h]
\centering
\centerline{ \includegraphics[width=0.4\textwidth]{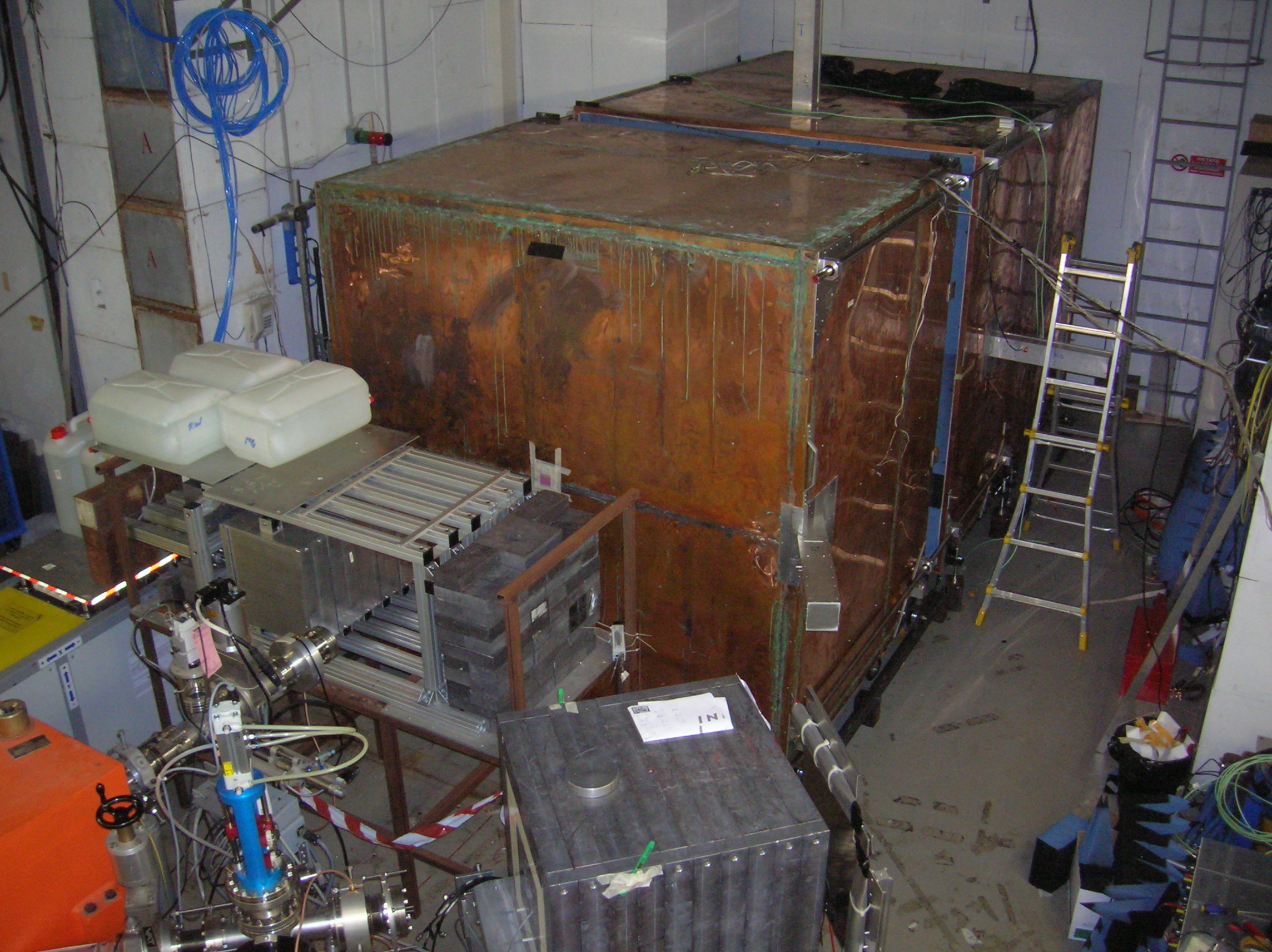} \hspace*{0.5cm} \includegraphics[width=0.35\textwidth]{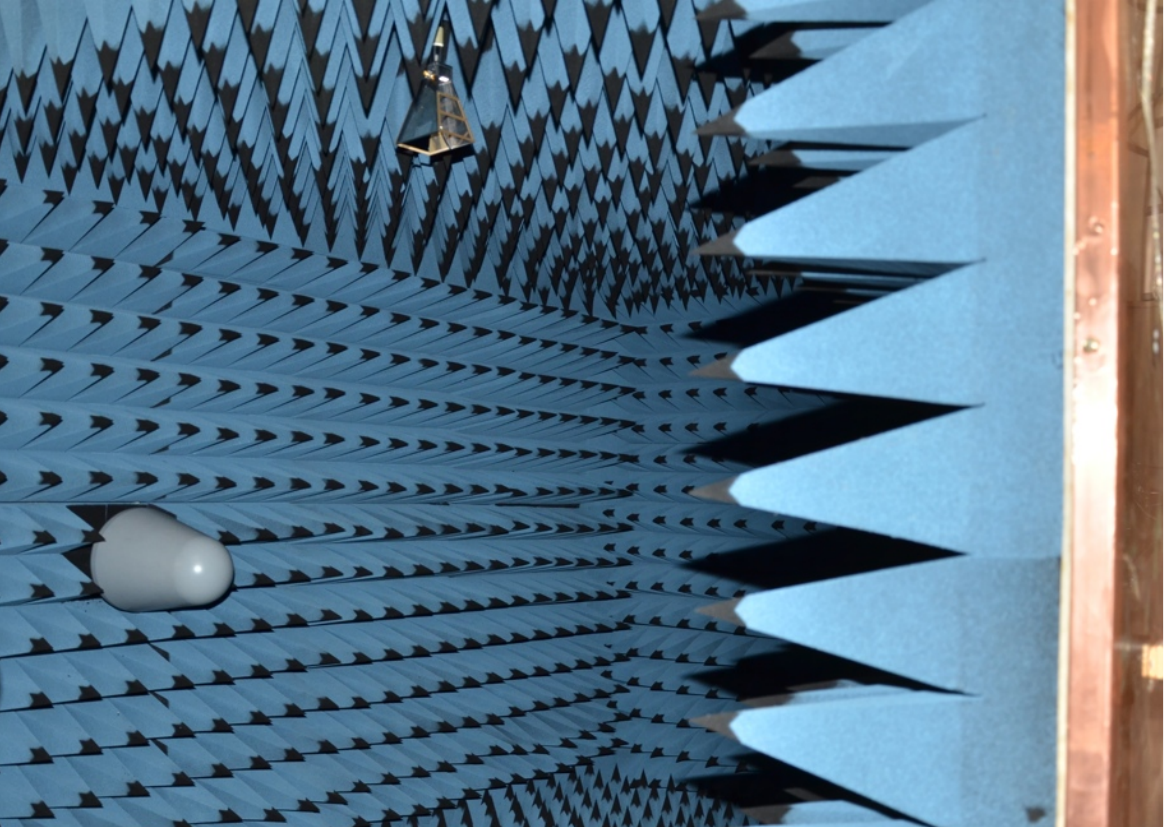} }
 \caption{Left panel: the AMY anechoic Faraday chamber installed at the BTF. Right panel: the antennas within the chamber.}
\label{Fig:ChamberAntennas}
\end{figure}
 transportation to the BTF, the chamber has been built in 3 modules: two have a length of 1.5 m and one is 1 m long. The left panel of Fig.~\ref{Fig:ChamberAntennas} shows the chamber installed at the BTF. The outer surface is covered by copper and connected to ground. The inner surface of the chamber (see right panel of Fig.~\ref{Fig:ChamberAntennas}) is covered by pyramidal RF absorbers (AEP-12 model), which have a height of 30 cm and an absorption range from 35 dB at 1 GHz to 45 dB at 6 GHz and 50 dB at higher frequencies. The shielding of the chamber from outside radiation is better than 80 dB above 2 GHz and it decreases quickly at lower frequencies (at 1 GHz it is $\sim$50 dB).

The AMY chamber may host the radio receivers in 5 different positions (corresponding to the aluminum boxes visible in Fig.~\ref{Fig:ChamberAntennas}). The support of the radio receivers allows us to change the distance of the antennas from the beam axis and to rotate their polarization plane.

\begin{figure}[h]
\centering
\centerline{ \includegraphics[width=0.37\textwidth]{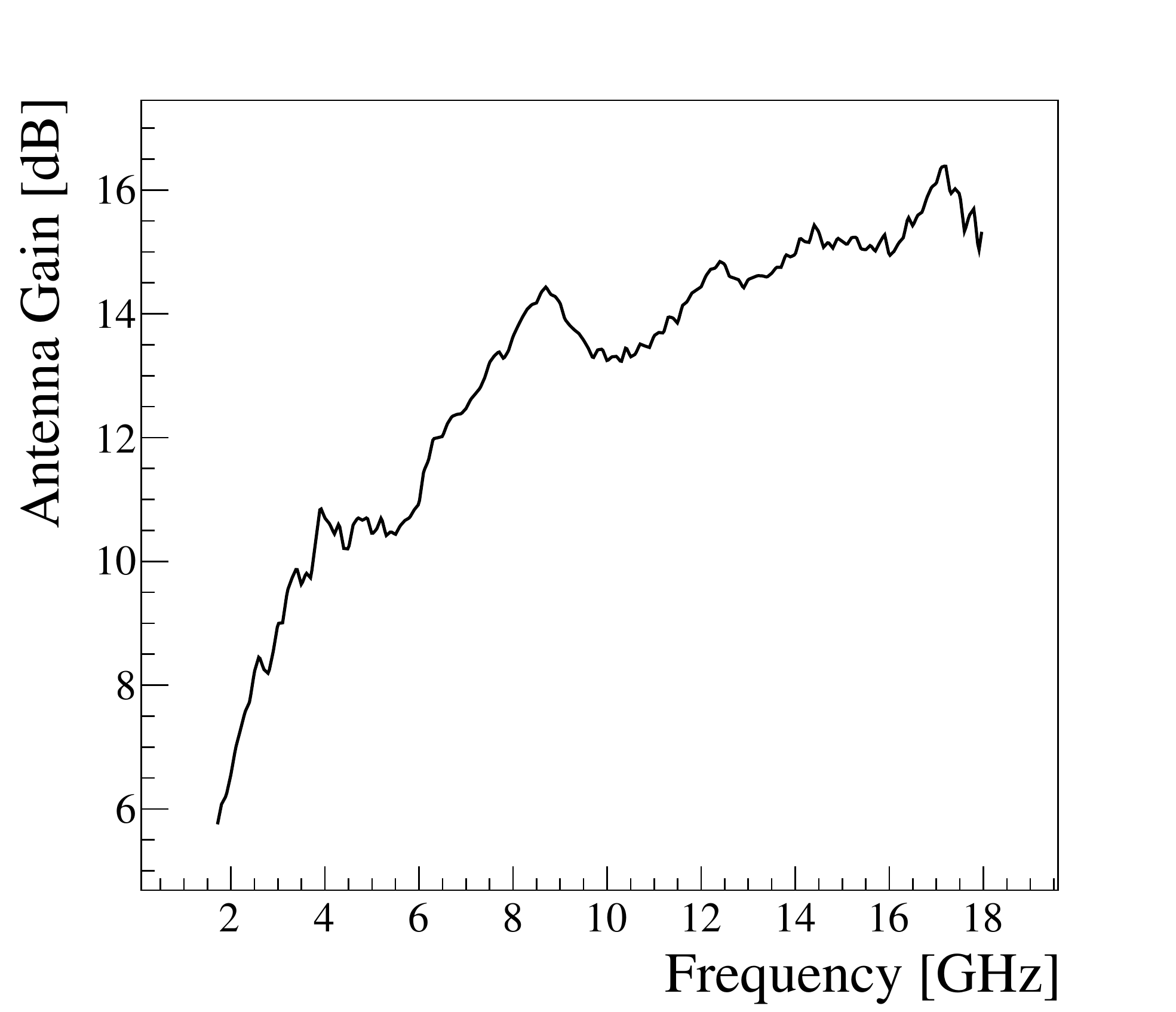} \includegraphics[width=0.37\textwidth]{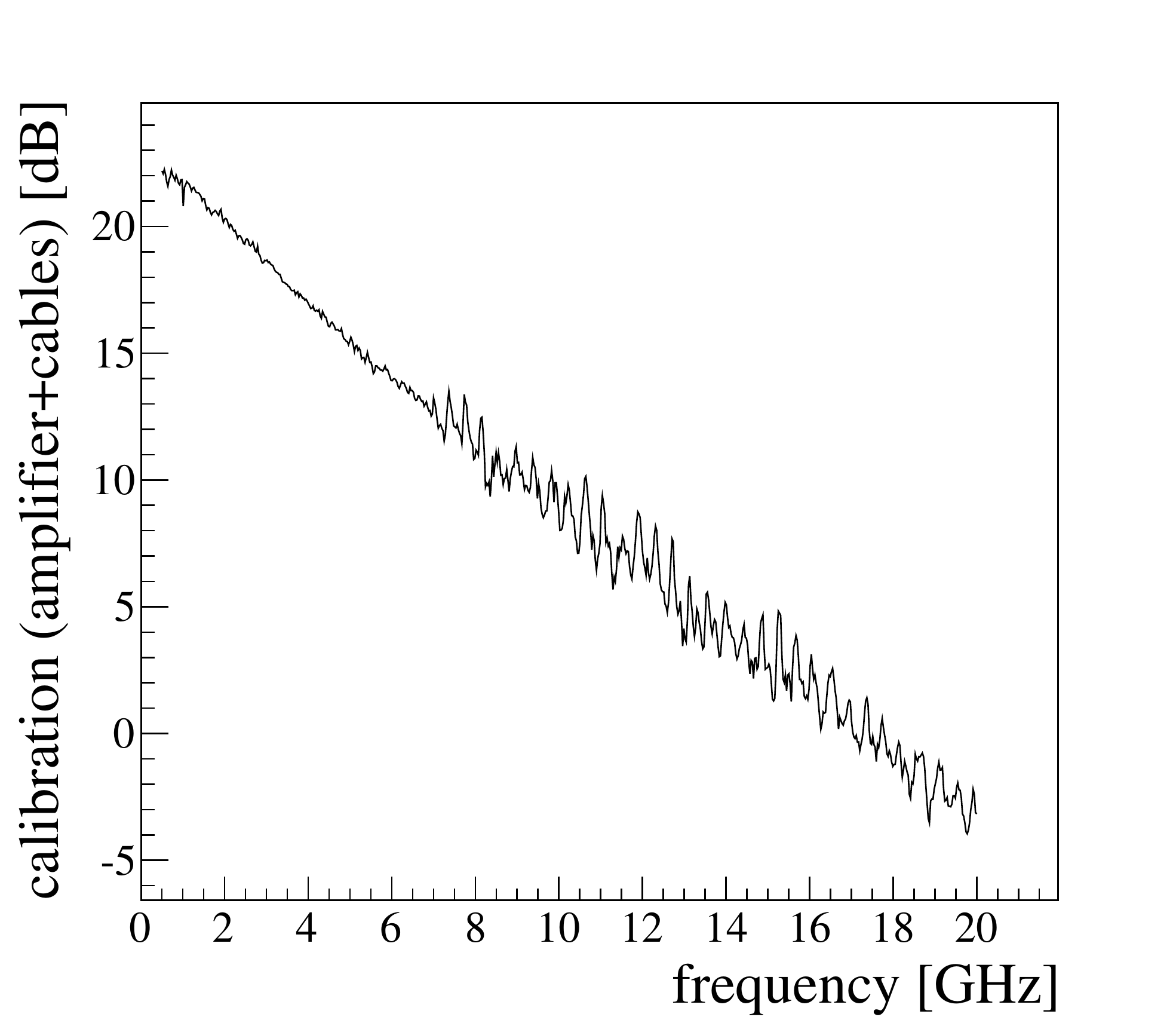} }
 \caption{Left panel: gain of the horn as a function of the frequency. Right panel: calibration factor that includes the amplifier and the loss in the cables.}
\label{Fig:Calibration}
\end{figure}

The receivers are two~\emph{Rohde\&Schwarz (R\&S) HL050} log-periodic antennas and two~\emph{RF Spin Double Ridged Waveguide Horn DRH20}. Both types of antennas are linearly polarized and operate in a broadband range of input frequencies, from about 1 to 20 GHz. The antennas have been calibrated at the SATIMO StarLab calibration system~\cite{satimo}. The gain of the horn as a function of the frequency is shown in the left panel of Fig.~\ref{Fig:Calibration}.

The output signal of each radio receiver was amplified by about 26 dB
using the ~\emph{Minicircuits} wide band amplifier~\emph{ZVA-213-S+}
(800 MHz$-$21 GHz). The signals were sent through $\simeq$ 20 m low
loss cables to the control room and acquired by a~\emph{Lecroy SDA
830Zi-A} oscilloscope, which has four input channels with a 20 GHz
real time bandwidth and a sampling rate of 40 GS/s. The channels were
calibrated using a~\emph{Rohde\&Schwarz SMF100A} signal generator
(range 100 kHz$-$22 GHz) and a~\emph{Rohde\&Schwarz SFSV30} spectrum
analyzer (range 9 kHz-30 GHz and 40 MHz bandwidth). The overall
calibration curve including the amplifier and cables is shown in the
right panel of Fig.~\ref{Fig:Calibration}. The loss at higher
frequencies is due to the cables.

\section{Data analysis}

The key point of the measurement is the understanding of high background arising from the Cherenkov and transition radiations produced by the relativistic electrons. These radiations are linearly polarized in the plane containing the beam axis and the direction of the radiation. 
\begin{figure}[h]
\centering
\centerline{ \includegraphics[width=0.3\textwidth]{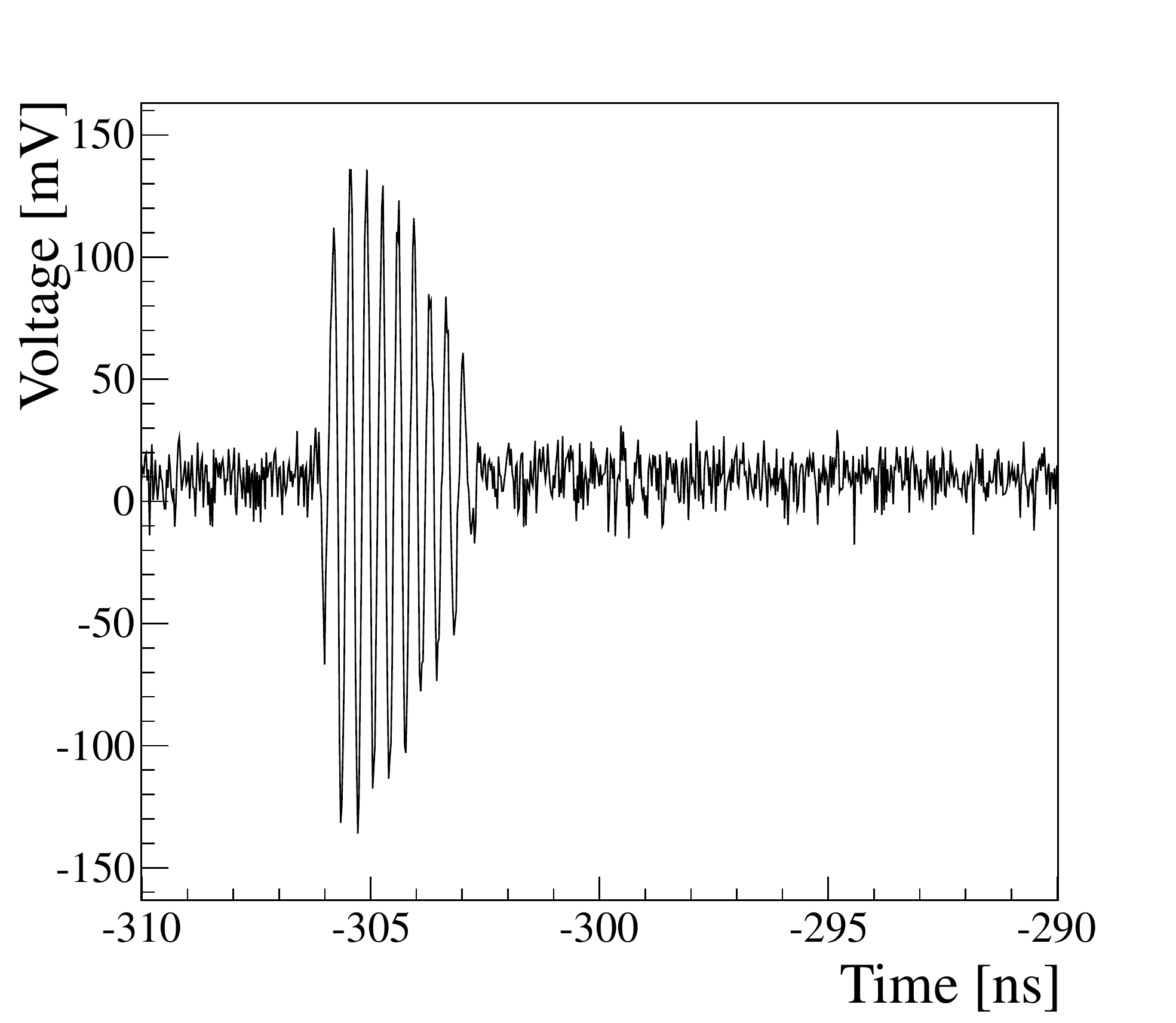} \includegraphics[width=0.3\textwidth]{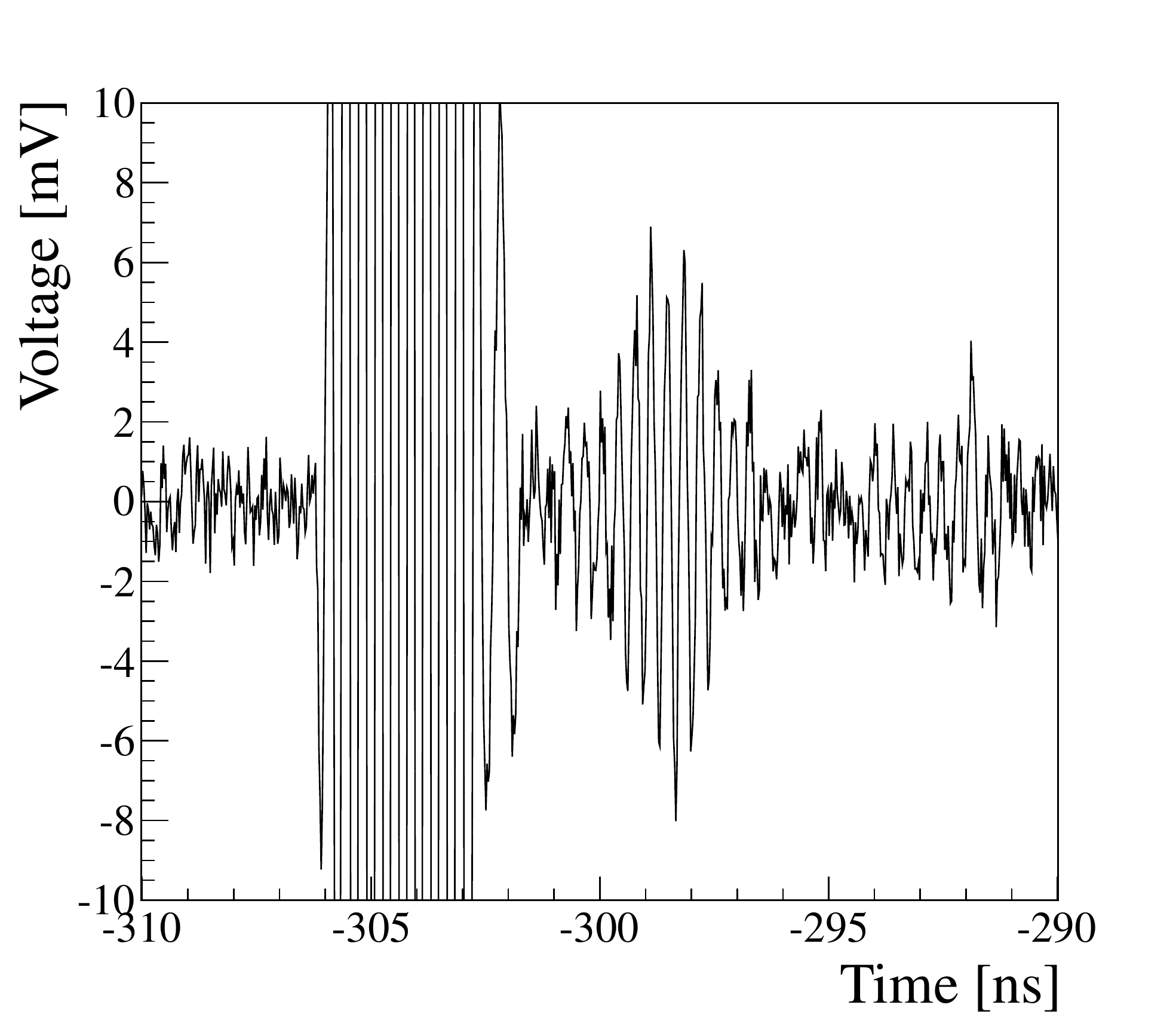} \includegraphics[width=0.3\textwidth]{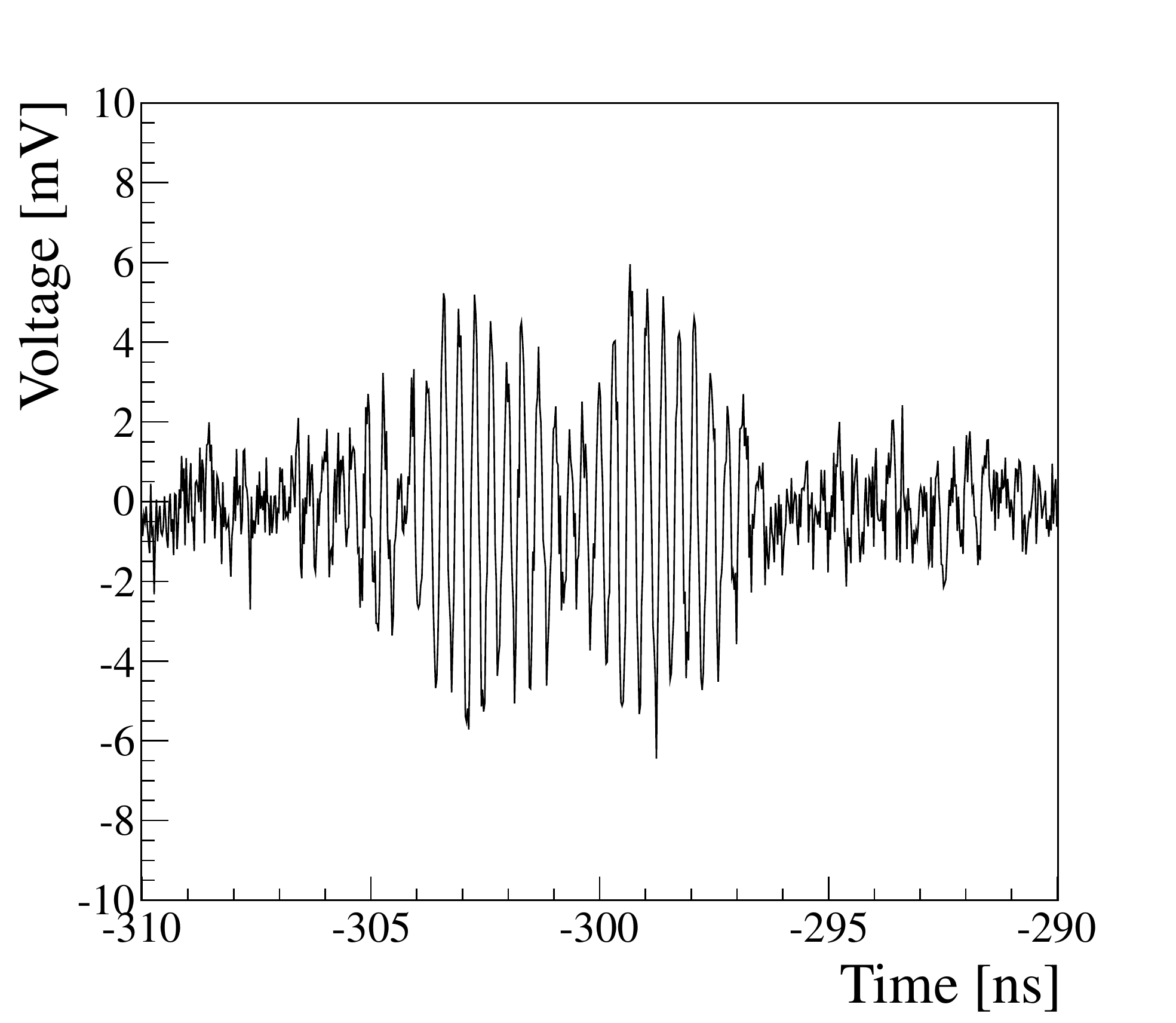} }
\caption{From the left to the right panel: co-pol signal, co-pol signal with oscilloscope chain saturated and cross-pol signal. In all three cases the beam intensity was the same.}
\label{Fig:CrossCoSignals}
\end{figure}
When the antenna polarization plane is parallel to the beam axis (co-pol configuration) the antenna signal is very high and dominated by the Cherenkov contribution as shown in the left panel of Fig.~\ref{Fig:CrossCoSignals}. When the polarization plane is perpendicular to the beam axis (cross-pol configuration) the signal is much lower as shown in the right panel of Fig.~\ref{Fig:CrossCoSignals}.

The signal has a double peak structure, clearly evident in the cross-pol configuration. We have made several tests that have excluded that this time structure is due to reflections in the electronic chain and in the chamber. Contrary to the first peak, the second one is un-polarized. This is shown in the central panel of Fig.~\ref{Fig:CrossCoSignals} where the co-pol signal of the left panel was acquired with the oscilloscope scale saturated, increasing in this way the sensitivity to low signals.

The second peak is very interesting because it is un-polarized, which is the experimental signature of the MBR that distinguishes it from the Cherenkov radiation. Moreover, it is slightly delayed in time and this is consistent with MBR and Cherenkov signals respectively dominated by the emission at the center and the beginning of the chamber.

\begin{figure}[h]
\centering
\centerline{ \includegraphics[width=0.35\textwidth]{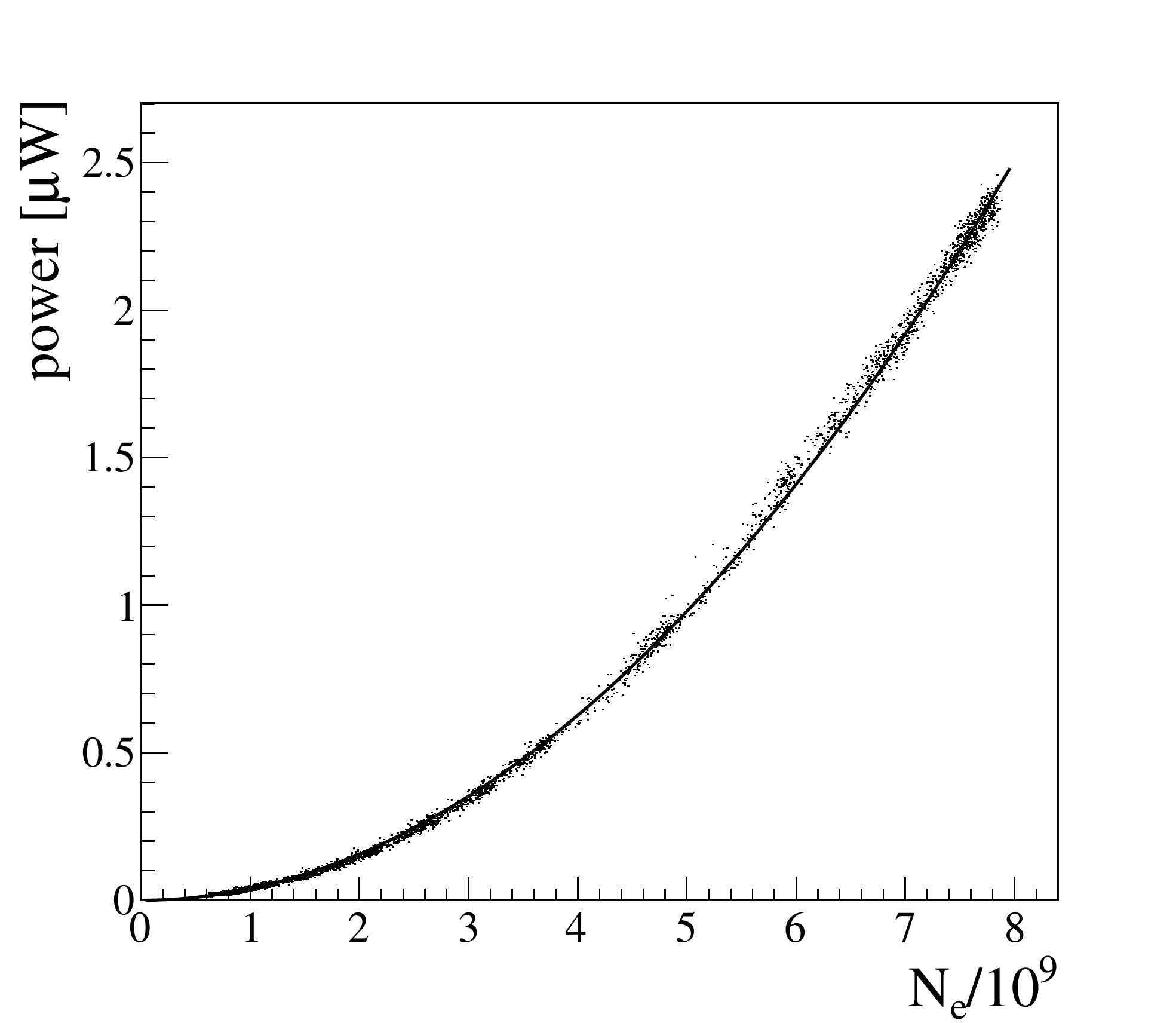} \includegraphics[width=0.35\textwidth]{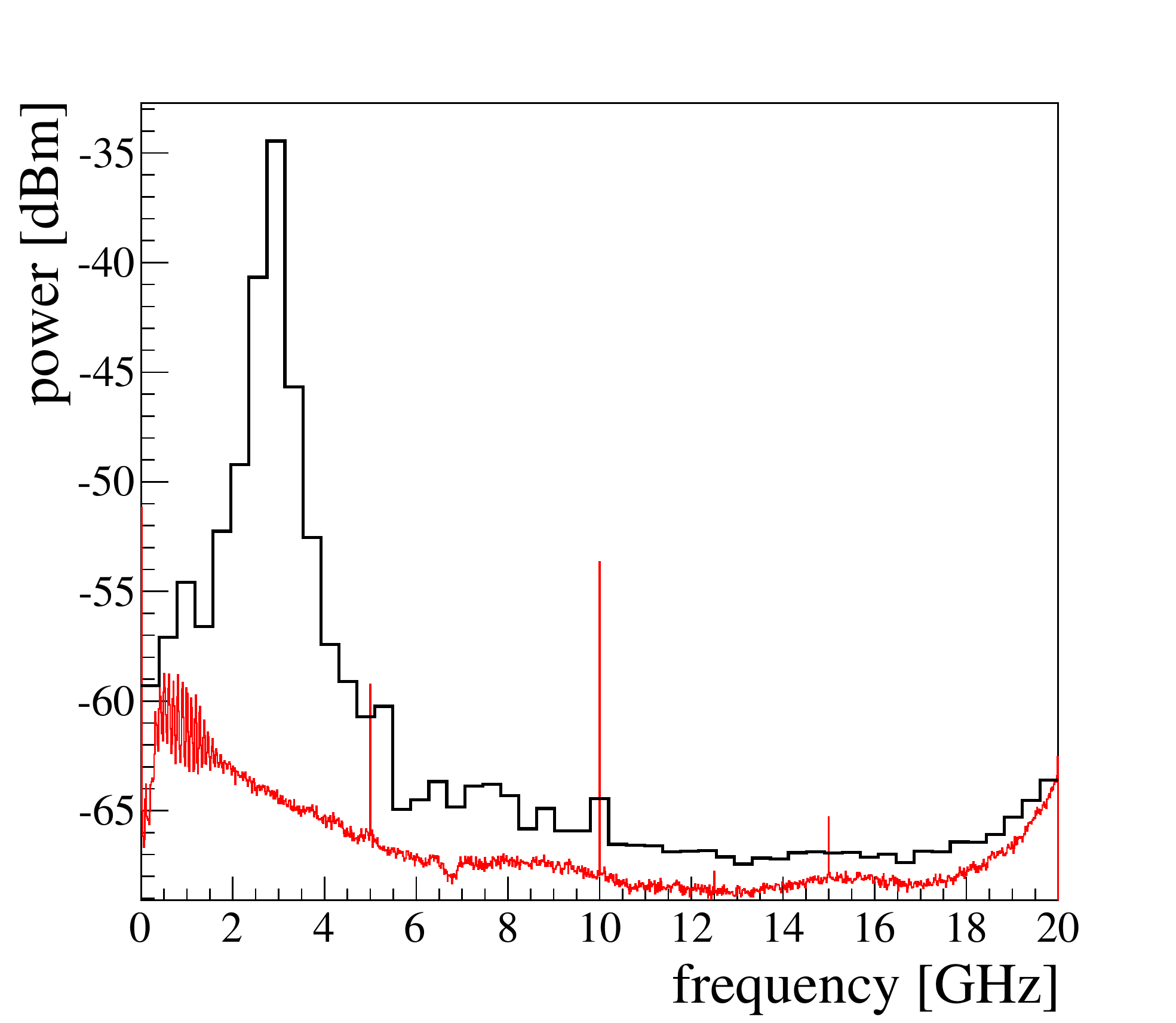} }
\caption{Left panel: power of the cross-pol signal as a function of the beam intensity. Right panel: frequency spectrum of the oscilloscope trace.}
\label{Fig:IntScanSpectrum}
\end{figure}
 
The cross-pol signal was found to be fully coherent. The power of the signal as a function of the number of electrons per bunch is well described by a quadratic power law as shown in the left panel of Fig.~\ref{Fig:IntScanSpectrum}. The power has been calculated from the root mean square of the oscilloscope trace in a time window around the signal and subtracting the background contribution calculated using the beginning of the trace, when the beam is not yet present in the chamber.

The right panel of Fig.~\ref{Fig:IntScanSpectrum} shows the frequency spectrum of the signal (black) obtained using the Fourier Transform of the oscilloscope traces and averaging over many events. The contribution from the background is shown in red. The spectrum is dominated by a peak centered at the LINAC frequency ($f_L$ = 2.856 GHz). The data refer to the cross-pol signal with a target of 2.3 X$_0$ thickness. The same analysis applied to the co-pol signal shows the same feature. In addition, the large intensity of the co-pol signal allows to point out the presence of other peaks at multiples of $f_L$.

We have developed a detailed simulation of the Cherenkov contribution according to the model developed in~\cite{jaime}. The simulation qualitatively reproduces the observed spectrum. The peaks of the signal are due to the constructive interference of the radiation emitted by the beam microbunches and the power of the signal scales quadratically with the beam intensity. Therefore we conclude that the coherence of the cross-pol signal is induced by the LINAC. This implies that the process generating the cross-pol signal is very prompt and we can not exclude that it could arise from the Cherenkov emission.

\begin{figure}[h]
\centering
\centerline{ \includegraphics[width=0.3\textwidth]{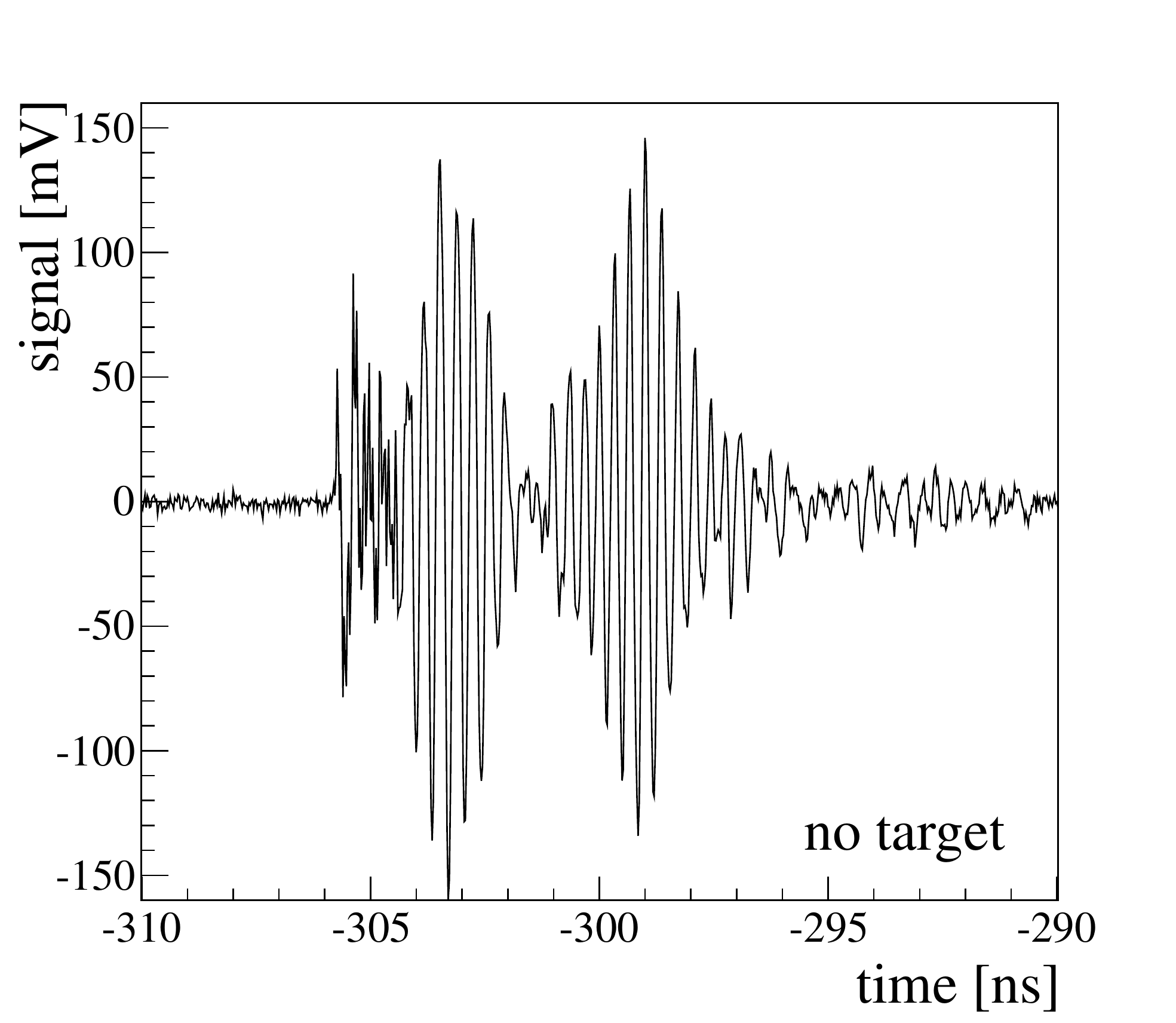} \includegraphics[width=0.3\textwidth]{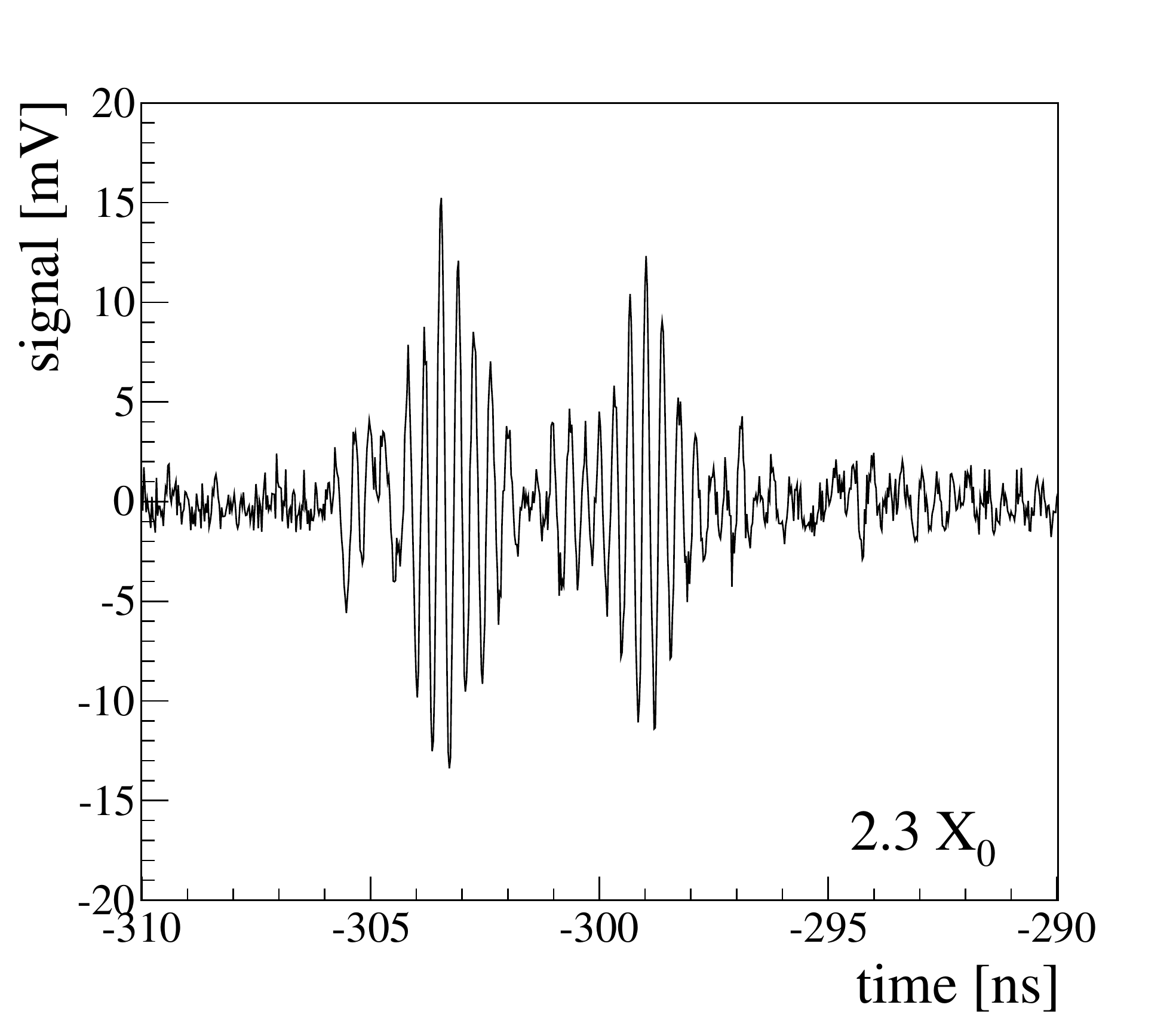} \includegraphics[width=0.3\textwidth]{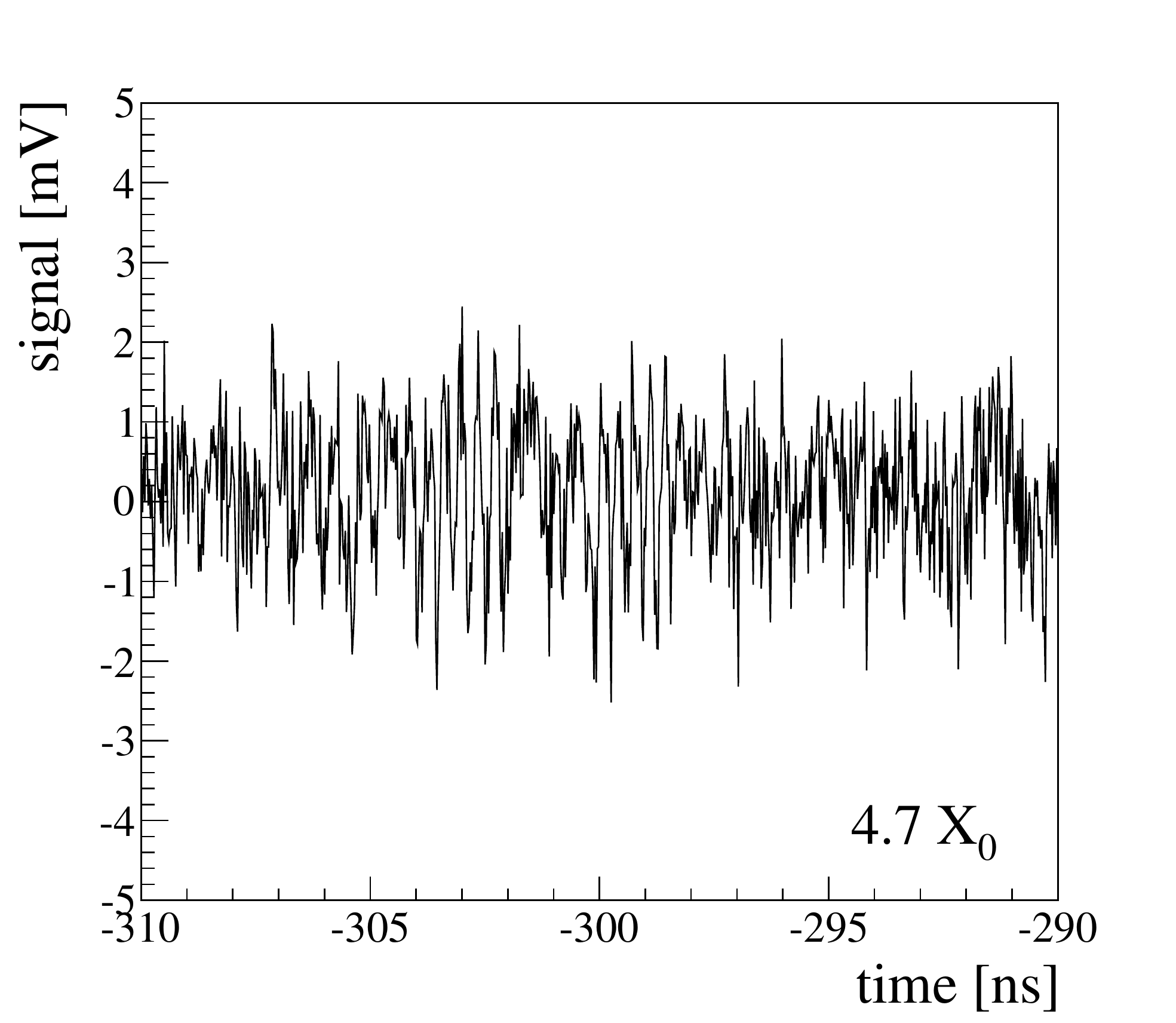} }
\caption{Cross-pol signal for different thicknesses of the interaction target.  The number of particles $N_e = 2.3 \times 10^9$ and the bunch length duration is 1.5 ns for all signals.}
\label{Fig:AntennaSignals}
\end{figure}

Another important feature of the cross-pol signal is the dependence between its intensity and the thickness of the interaction target. Fig.~\ref{Fig:AntennaSignals} shows the raw oscilloscope traces for $N_e = 2.3 \times 10^9$ when there is no target and for 2.3 and 4.7 X$_0$. The signal decreases from $\sim$100 mV to $\sim$1 mV which corresponds to a factor $10^4$ in power. This fall of the power can not be explained by the change of the energy deposit, which is only a factor 5 (see Fig.~\ref{Fig:BunchTarget}). A similar trend has been observed for the co-pol signal. This behavior is interpreted as the result of a loss of coherence, which is caused by a larger average distance between the electrons after their collision with the target.

\section{Conclusions}
We have presented the AMY experiment and the first results of the
measurements done at BTF, Frascati. We do not confirm the previous
results obtained at SLAC~\cite{Gorham-SLAC}. The observed signal seems
to be very prompt like the Cherenkov radiation and the coherence is
certainly caused by the particular time structure of the beam. The AMY
measurements can be used to put an upper limit on the intensity of the
MBR process. The best limit can be obtained with the largest target
thickness. The analysis is still preliminary. Using the signal with
4.7 X$_0$, the calibration of the detector and considering that the
signal is mainly concentrated in a narrow bandwidth of about 1 GHz
around the LINAC frequency, we obtain a density flux $< 10^{-16}$
W/m$^2$/Hz.


\begin{theacknowledgments}
   The work is supported 
     by INFN, Italy, 
    the EC FP7 Research Infrastructure projects (HadronPhysics3, GA n. 283286 and AIDA, GA 262025), 
    the MSMT CR grant Lg13007 
    and by the 2$^{\rm nd}$ ASPERA Call (7$^{\rm nd}$ EU ERA-NET program).
\end{theacknowledgments}



\bibliographystyle{aipproc}   




\end{document}